# Relaxing Direct Ptychography Sampling Requirements via Parallax Imaging Insights

Georgios Varnavides[1,2,*], Julie Marie Bekkevold[3,*], Stephanie M Ribet[2], Mary C Scott[2,4], Lewys Jones[3], Colin Ophus[5]

[1] Miller Institute for Basic Research in Science, University of California Berkeley, Berkeley, USA
[2] National Center for Electron Microscopy, Lawrence Berkeley National Laboratory, Berkeley, USA
[3] School of Physics, Trinity College Dublin, Dublin, Ireland
[4] Department of Materials Science and Engineering, University of California Berkeley, Berkeley, USA
[5] Department of Materials Science and Engineering, Stanford University, Stanford, USA

* These authors contributed equally to this work
Correspondence may be addressed to: Georgios Varnavides, gvarnavides@berkeley.edu; Colin Ophus, cophus@gmail.com

## Abstract

Direct ptychography enables the retrieval of information encoded in the phase of an electron wave passing through a thin sample by deconvolving the interference effects of a converged probe with known aberrations. Under the weak phase object approximation, this permits the optimal transfer of information using non-iterative techniques. However, the achievable resolution of the technique is traditionally limited by the probe step size – setting stringent Nyquist sampling requirements. At the same time, parallax imaging has emerged as a dose-efficient phase-retrieval technique which relaxes sampling requirements and enables scan-upsampling. Here, we formulate parallax imaging as a quadratic approximation to part of the direct ptychography kernel and use this insight to enable upsampling in direct ptychography. We validate our analytical results numerically using simulated and experimental reconstructions.

## Introduction

Phase retrieval techniques in scanning transmission electron microscopy (STEM) are increasingly growing in popularity due to their unrivalled dose efficiency and spatial resolution, making them well-suited for various materials applications (Sanchez-Santolino et al., 2025, Varnavides et al., 2023). This has in part been driven by hardware developments in fast direct electron detectors (Levin, 2021), allowing the acquisition of the diffracted probe intensity at each converged probe position – forming a broader family of techniques known as 4D-STEM (Ophus, 2019). Among 4D-STEM phase retrieval techniques, ptychography attempts to retrieve the phase imparted by a thin sample on the incoming illumination by deconvolving the converged probe from a set of measured intensities.

Ptychography is often differentiated between 'direct' methods, such as single side-band (SSB) (Pennycook et al., 2015; Yang et al., 2016a, 2015), and Wigner distribution deconvolution (WDD) (Rodenburg and Bates, 1992, Yang et al., 2017), and 'iterative' methods which refers to a variety of model-based reconstruction algorithms (Chen et al., 2020, Rodenburg and Maiden, 2019). While direct ptychography is computationally much easier to perform and, for weak phase objects, can be shown analytically to recover the sample information accurately (Nellist et al., 1995), three reasons are often cited for iterative ptychography's recent prevalence in the literature:

1. 'super resolution', i.e. recovering information beyond the diffraction limit given by the numerical aperture (twice the convergence semiangle),
2. superior depth slicing resolution, enabled using multislice forward and adjoint models, and
3. relaxed sampling requirements, enabled by defocused acquisitions with large step sizes.

Parallax imaging, or tilt-corrected bright field (tcBF) STEM (Spoth et al., 2017; Varnavides et al., 2024, 2023; Yu et al., 2024), is a relatively new phase retrieval technique which relaxes sampling requirements by providing post-acquisition upsampling of the reconstructed phase. This results in the ability to recover information well beyond the Nyquist limit implied by the scan step size. However, parallax imaging does not recover sample information as accurately as direct ptychography – exhibiting zero-crossings and contrast reversals (Bekkevold et al., 2025, Ma et al., 2025, Yu et al., 2024).

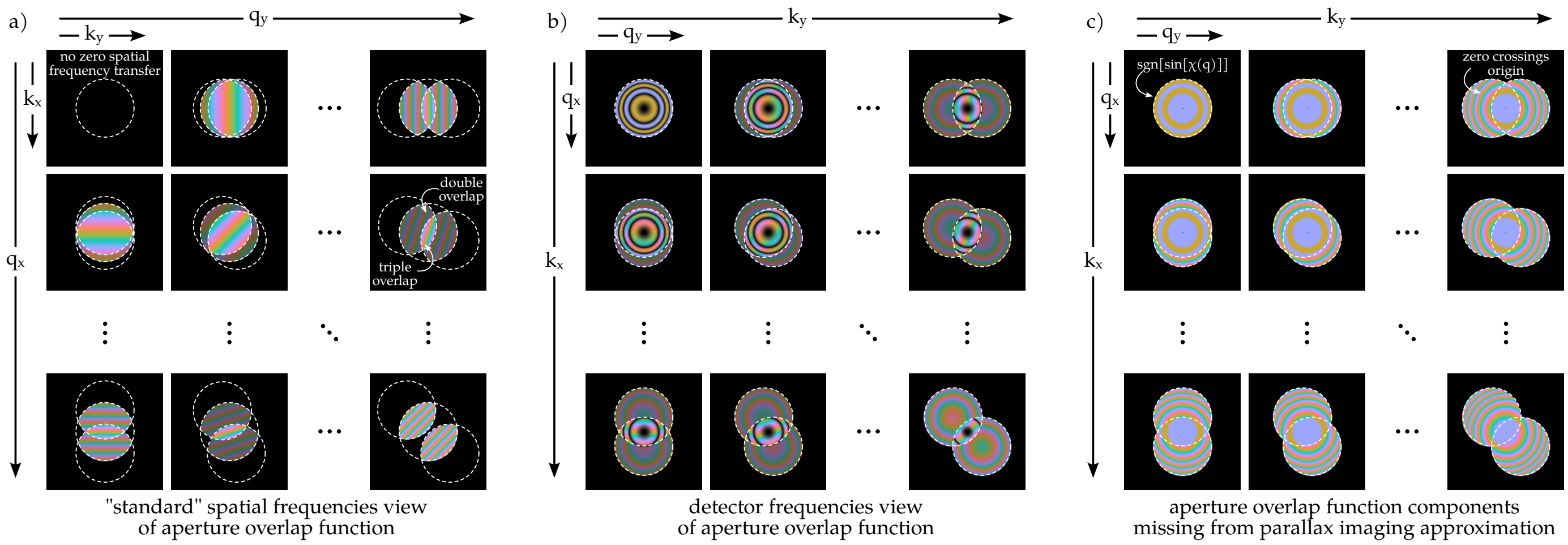

Figure 1: **Different views of the complex-valued aperture overlap function,** $\Gamma(\boldsymbol{q},\boldsymbol{k})$. **a)** Fixing specific spatial frequencies $\boldsymbol{q}$ and plotting over the detector frequencies $\boldsymbol{k}$, illustrates the so-called 'double' and 'triple' overlap 'trotter' regions. **b)** Fixing specific detector frequencies $\boldsymbol{k}$ and plotting over the spatial frequencies $\boldsymbol{q}$, illustrates the effect of shifting the probe aperture and aberrations. **c)** Plots the part of the $\Gamma(\boldsymbol{q},\boldsymbol{k})$ function not captured by parallax imaging, demonstrating the necessary $\mathrm{sgn}[\sin[\chi(\boldsymbol{q})]]$ phase-flipping in the $\boldsymbol{k}=0$ subpanel, and the origin of the zero-crossings.

In this work, we elucidate the link between parallax imaging and direct ptychography, and use this insight to propose an upsampling scheme for direct ptychography. This relaxes the stringent sampling requirements of direct ptychography, enabling its application to dose-sensitive samples such as organic solids and biological proteins.

The manuscript is structured as follows. First, we present the transfer of information formalism for direct ptychography and parallax imaging, following Hammel and Rose (1995). We use this to show how parallax imaging can be obtained as a truncated Taylor series expansion of the complex-valued direct ptychography kernel, and demonstrate an upsampling scheme for direct ptychography. We investigate the limits of this upsampling scheme numerically, highlighting its dependence on the magnitude of the aberration surface. Finally, we perform upsampled direct ptychography reconstructions on simulated and experimental datasets to demonstrate their utility.

## Transfer of Information Formalism

In general, STEM image formation theory is non-linear. For weak-phase objects however, the sample potential is approximately proportional to the induced phase shift (Hammel and Rose, 1995). This allows us to define a sample-independent contrast transfer function (CTF) which depends solely on the details of the imaging system, such as the incoming illumination, the detector geometry, and the reconstruction algorithm. This is commonly expressed in terms of the reconstructed image Fourier transform:

$$\begin{aligned}\tilde{I}_j(\boldsymbol{q}) &= 2\tilde{\varphi}(\boldsymbol{q}) \times \mathcal{L}_j(\boldsymbol{q}) \\ \tilde{I}_j(\boldsymbol{q}) &= \mathcal{F}_{\boldsymbol{r}\rightarrow\boldsymbol{q}}\big[I_j(\boldsymbol{r})\big],\end{aligned} \tag{1}$$

where $\tilde{\varphi}(\boldsymbol{q})$ and $\tilde{I}_j(\boldsymbol{q})$ are the Fourier sample phase and reconstructed image, evaluated at spatial frequency $\boldsymbol{q}$. The complex-valued CTF for the $j^{\text{th}}$ detector segment, $\mathcal{L}_j(\boldsymbol{q})$, is given by (Bekkevold et al., 2025, Hammel and Rose, 1995):

$$\begin{aligned}\mathcal{L}_j(\boldsymbol{q}) = \frac{i}{2}\int A(\boldsymbol{k})D_j(\boldsymbol{k})\big\{&A(\boldsymbol{q}-\boldsymbol{k})\mathrm{e}^{-\mathrm{i}[\chi(\boldsymbol{q}-\boldsymbol{k})-\chi(\boldsymbol{k})]} \\ &-A(\boldsymbol{q}+\boldsymbol{k})\mathrm{e}^{\mathrm{i}[\chi(\boldsymbol{q}+\boldsymbol{k}-\chi(\boldsymbol{k}))]}\big\},\end{aligned} \tag{2}$$

where $A(\boldsymbol{k})$ and $\chi(\boldsymbol{k})$ are the probe-forming aperture and aberration surface respectively, evaluated at detector frequency, $\boldsymbol{k}$. $A(\boldsymbol{k})$ is a normalized top-hat function and

$$\chi(k,\theta) = \frac{2\pi}{\lambda}\sum_{n,m}\frac{1}{n+1}C_{n,m}(k\lambda)^{n+1}\cos\big[m(\theta-\theta_{n,m})\big] \tag{3}$$

where $k=|\boldsymbol{k}|$, $\theta=\arctan[\boldsymbol{k}]$, $\lambda$ is the relativistic electron wavelength, $n$ and $m$ are radial and azimuthal orders of the aberration coefficients $C_{n,m}$ with aberration axis $\theta_{n,m}$. For a pixelated detector[i], $D_j(\boldsymbol{k})=\delta(\boldsymbol{k})$, the integrand of Equation 2 reduces to (Ma et al., 2025, Yu et al., 2024):

$$\begin{aligned}\Gamma(\boldsymbol{q},\boldsymbol{k}) = &A(\boldsymbol{k})A(\boldsymbol{q}-\boldsymbol{k})\mathrm{e}^{-\mathrm{i}\chi(\boldsymbol{q}-\boldsymbol{k})}\mathrm{e}^{\mathrm{i}\chi(\boldsymbol{k})} \\ &-A(\boldsymbol{k})A(\boldsymbol{q}+\boldsymbol{k})\mathrm{e}^{\mathrm{i}\chi(\boldsymbol{q}+\boldsymbol{k})}\mathrm{e}^{-\mathrm{i}\chi(\boldsymbol{k})}\end{aligned} \tag{4}$$

This is known as the complex-valued aperture overlap function (Yang et al., 2016a), usually written compactly as:

$$\begin{aligned}\Gamma(\boldsymbol{q},\boldsymbol{k}) &= \psi^*(\boldsymbol{k})\psi(\boldsymbol{q}-\boldsymbol{k}) - \psi(\boldsymbol{k})\psi^*(\boldsymbol{q}+\boldsymbol{k}) \\ \psi(\boldsymbol{k}) &= A(\boldsymbol{k})\mathrm{e}^{-\mathrm{i}\chi(\boldsymbol{k})}\end{aligned} \tag{5}$$

### Direct Ptychography

Figure 1(a) plots Equation 5 over the detector frequencies $\boldsymbol{k}$ for specific values of spatial frequencies $\boldsymbol{q}$, for a defo-cused probe. Note that there is no transfer of information at frequency $\boldsymbol{q} = 0$; this is true for all in-line holographic techniques used for STEM phase retrieval. For nonzero spatial frequencies, $\Gamma(\boldsymbol{q}, \boldsymbol{k})$ exhibits the so-called double and triple overlap regions or 'trotters' (Clark and Nellist, 2025, Nellist and Rodenburg, 1998, Rodenburg, 1989).

[i]Direct ptychography and parallax imaging can also be performed using segmented detectors, see Bekkevold et al. (2025) for a review.

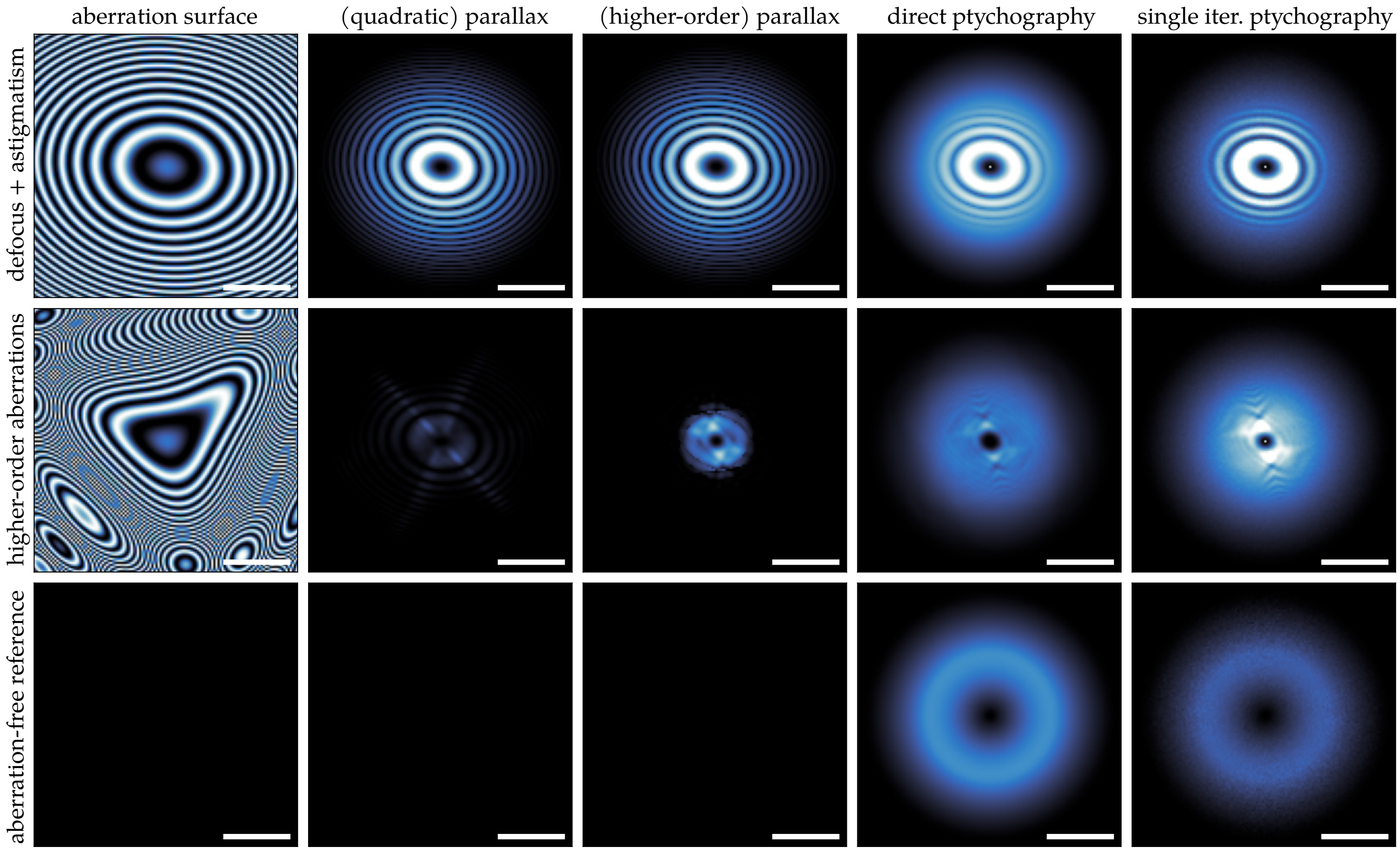

Figure 2: **Contrast transfer of information for direct ptychography and parallax imaging.** For quadratic aberrations, such as defocus and astigmatism (top row), parallax closely resembles direct ptychography, albeit including zero-crossings. In the present of higher-order aberrations, traditional parallax fails qualitatively to reconstruct the sample phase. Including the aberration surface Hessian (middle column), recovers some of the missing information, highlighting that parallax imaging can be thought-of as a truncated Taylor expansion for direct ptychography. The last column illustrates that direct ptychography itself can be thought-of as an approximation to iterative ptychography, obtained from a single iteration. Scalebars equal to $k_0$.

Modern direct ptychographic implementations are able to utilize all the phase information contained in the aperture overlap function (Yang et al., 2016a), and thus their CTF is given by summing the non-zero regions of Equation 5:

$$\mathcal{L}^{\text{ptycho}}(\boldsymbol{q}) = \frac{i}{2} \int | \, \Gamma(\boldsymbol{q}, \boldsymbol{k})| \; d\boldsymbol{k}. \tag{6}$$

It is instructive to plot $\Gamma(\boldsymbol{q}, \boldsymbol{k})$ over the spatial frequencies $\boldsymbol{q}$ for specific detector frequencies $\boldsymbol{k}$, shown in Figure 1(b). In this way, the conjugate relationship between the positively and negatively shifted apertures becomes apparent.

Note that, due to the common $A(\boldsymbol{k})$ term in Equation 4, this only has support over detector frequencies inside the bright-field (BF) disk, $\boldsymbol{k}_{\text{BF}} = \{\boldsymbol{k} : |\boldsymbol{k}| < k_0\}$. This suggests that, instead of the usual loop over spatial frequencies $\boldsymbol{q}$, we can equivalently loop over the BF detector frequencies, $\boldsymbol{k}_{\text{BF}}$, multiplying the Fourier-transformed virtual BF (vBF) images with the filters shown in Figure 1(b).

## Parallax Imaging

While intuitive, the way parallax imaging is often introduced – through the principle of reciprocity and vBF image shifts (Ma et al., 2025; Varnavides et al., 2024, 2023; Yu et al., 2024) – obscures its close connection with direct ptychography. More formally, parallax imaging can be thought of as multiplying the Fourier-transformed vBF images with the phase ramp $\mathrm{e}^{\mathrm{i}\nabla\chi(\mathbf{q})\cdot\mathbf{k}}$. By the Fourier-shift theorem, this is equivalent to shifting the vBF images by the gradient of the aberration surface, $\nabla\chi(\boldsymbol{q})$.

We can thus derive parallax imaging from direct ptychography by factoring out the phase ramp contribution from the aperture overlap function (Bekkevold et al., 2025):

$$\begin{aligned} \Gamma(\boldsymbol{q}, \boldsymbol{k}) &\approx \mathrm{B}(\boldsymbol{q}, \boldsymbol{k}) \mathrm{e}^{\mathrm{i}\nabla\chi(\mathbf{q})\cdot\mathbf{k}} \\ \mathrm{B}(\boldsymbol{q}, \boldsymbol{k}) &= A(\boldsymbol{k}) \left[ A(\boldsymbol{q} - \boldsymbol{k}) \mathrm{e}^{-\mathrm{i}\chi(\boldsymbol{q})} - A(\boldsymbol{q} + \boldsymbol{k}) \mathrm{e}^{\mathrm{i}\chi(\boldsymbol{q})} \right] \end{aligned} \tag{7}$$

Figure 1(c) plots $\mathrm{B}(\boldsymbol{q}, \boldsymbol{k})$ over the spatial frequencies $\boldsymbol{q}$ for specific detector frequencies $\boldsymbol{k}$. Looking at the $\boldsymbol{k} = 0$ subpanel, it is interesting to note that direct ptychography implicitly performs the $\mathrm{sgn}[\sin[\chi(\boldsymbol{q})]]$ phase-flipping often performed in post-processing for parallax imaging. This connection is also implicitly discussed in Ma et al. (2025).

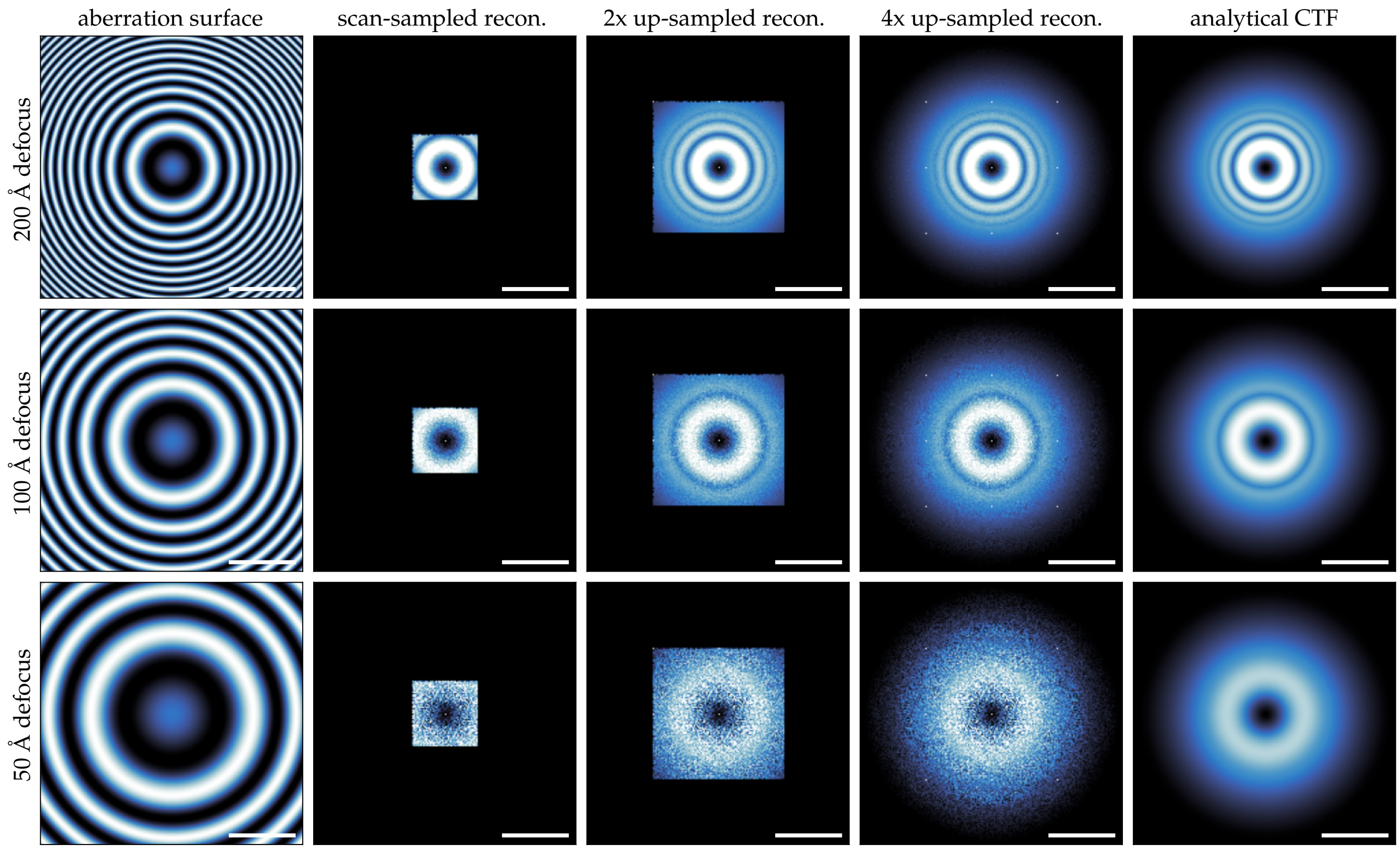

Figure 3: **Upsampling limits in direct ptychography.** Subsampled direct ptychography acquisitions (second column), upsampled to 2x (third column) and 4x (fourth column) scan resolution. The last column plots the analytical CTF the upsampled reconstructions are attempting to recover. Each row represents an increasing amount of defocus, highlighting upsampling factor limits are proportional to aberrations magnitude. Scalebars equal to $k_0$.

The parallax CTF is obtained by summing $\mathrm{B}(\boldsymbol{q},\boldsymbol{k})$ over $\boldsymbol{k}$:

$$\begin{aligned}\mathcal{L}^{\mathrm{prlx}}(\boldsymbol{q}) &= \frac{i}{2}\int \mathrm{B}(\boldsymbol{q},\boldsymbol{k})d\boldsymbol{k} \\ &= -[A\star A](\boldsymbol{q})\sin[\chi(\boldsymbol{q})],\end{aligned} \tag{8}$$

where $\star$ denotes cross-correlation. We recognize Equation 8 as the aperture autocorrelation function, modulated by the high-resolution TEM CTF, $-\sin[\chi(\boldsymbol{q})]$.

Figure 2 plots the direct ptychography and parallax imaging CTFs for various probe aberrations. For probes dominated by quadratic aberrations (Figure 2, top row), such as defocus and astigmatism, the parallax imaging CTF closely resembles that of direct ptychography – albeit including zero-crossings. However, for probes dominated by higher-order aberrations (Figure 2, middle row), such as coma and spherical aberration, parallax imaging fails qualitatively with a vanishing CTF.

This can be understood by noticing that Equation 7 is a Taylor series expansion of the shifted aberration surface:

$$\chi(\boldsymbol{q}\pm\boldsymbol{k}) \approx \chi(\boldsymbol{q}) \pm \nabla\chi(\boldsymbol{q})^T\cdot\boldsymbol{k} + \frac{1}{2}\boldsymbol{k}^T\cdot\mathcal{H}(\boldsymbol{q})\cdot\boldsymbol{k} + ... \tag{9}$$

where $\mathcal{H}(\boldsymbol{q})$ is the Hessian of the aberration surface, and parallax imaging truncates Equation 9 after the first order. It's interesting to note that while truncating Equation 9 to first order implies Equation 7 is only accurate to aberration terms linear in $\boldsymbol{q}$, the Hessian term for quadratic aberrations is constant and thus the second-order error term introduced by omitting this is independent of $\boldsymbol{q}$, contributing a global, physically irrelevant, phase shift.

The middle column of Figure 2 highlights how modifying parallax to include the Hessian term recovers some of the missing information, but is still inferior to direct ptychography. For completeness, the last column of Figure 2 completes this approximation ladder by highlighting that the direct ptychography CTF can be obtained using a single iteration of iterative ptychography. The bottom row highlights that, in contrast to parallax, ptychography does not require aberrations to transfer sample information.

## Relaxing Sampling Requirements

As discussed in the introduction, one of the benefits of iterative ptychography and parallax imaging is that the achievable resolution is largely decoupled from the scan step size (Varnavides et al., 2023). In iterative ptychography, the maximum achievable resolution is set by the largest scattering angle recorded with meaningful scattering (Rodenburg and Maiden, 2019). In parallax imaging, reconstructions can be upsampled post-acquisition (Nguyen et al., 2016, Spoth et al., 2017, Yu et al., 2022), to a maximum achievable resolution set by the numerical aperture (Varnavides et al., 2024, 2023; Yu et al., 2024).

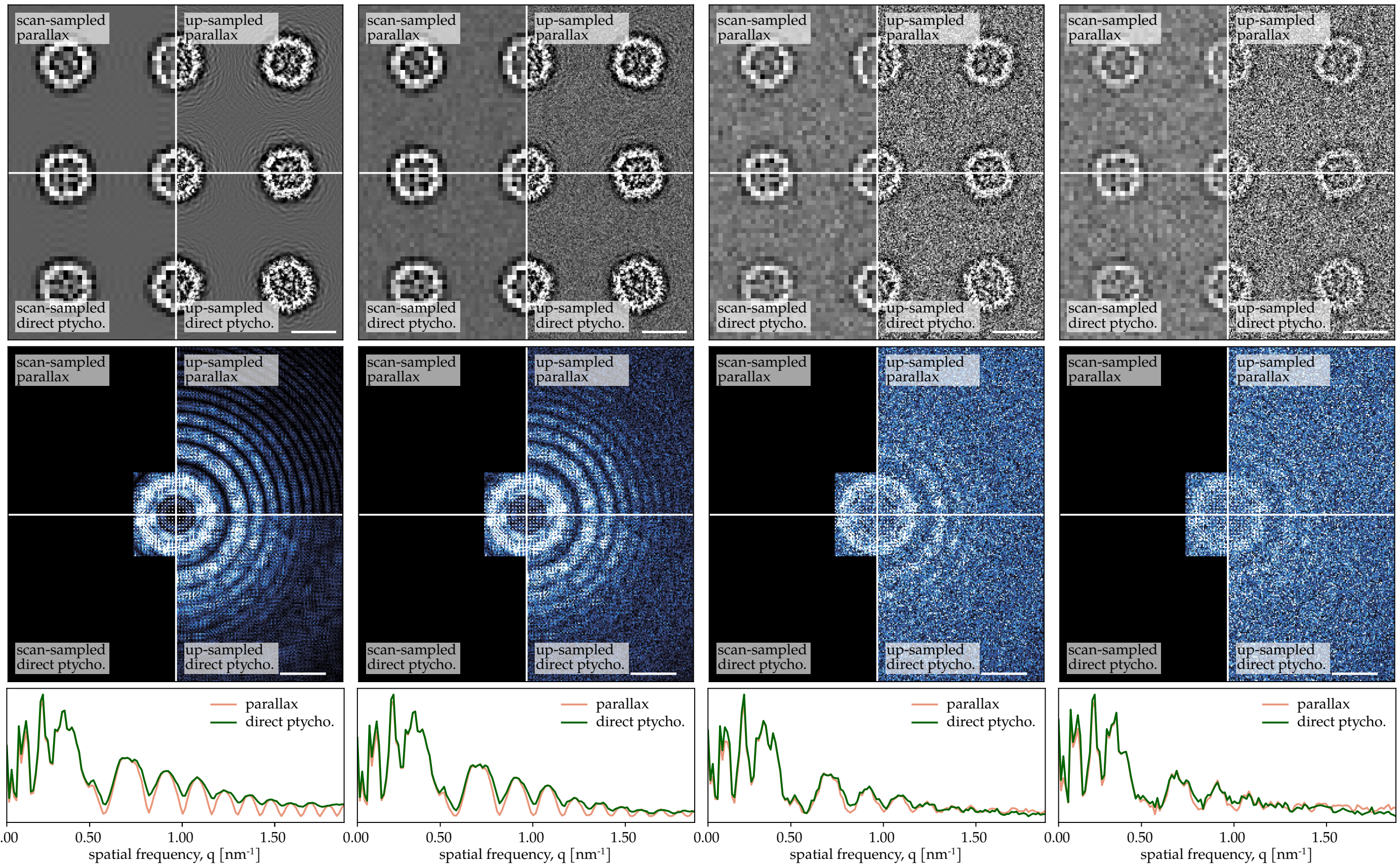

Figure 4: **Zero crossing recovery limits.** Parallax imaging (top half) and direct ptychography (bottom half) reconstructions for simulated apoferritin proteins, and decreasing electron fluence from left to right. At finite electron fluence, the recovery of the CTF zero crossings enabled by direct ptychography become less important – highlighting the utility of parallax imaging for low dose imaging. Scalebars are 10 nm and 0.5 $\text{nm}^{-1}$ respectively.

Traditionally, the resolution in direct ptychography is set by the Nyquist limit of the scan step size, with the maximum achievable resolution again set by the numerical aperture (Pennycook et al., 2015). Recently, Lalandec Robert et al. (2025) proposed an algorithm to relax the scan sampling requirements in direct ptychography by explicit Einstein summations over the scan frequencies. In this work, we propose an alternative algorithm to relax scan sampling requirements by reformulating direct ptychography as a loop of BF detector frequencies, instead of scan frequencies. This ensures we can still use efficient numerical implementations like the fast Fourier transform. Algorithm 1 sketches our upsampling implementation, which shares many similarities to parallax imaging, albeit using the full aperture overlap kernel in Equation 4.

Figure 3 demonstrates the upsampled CTF in direct ptychography, highlighting the upsampling limits and their dependence on the magnitude of the aberrations. First, note that in the presence of 'sufficient' aberrations (top

---

**Algorithm 1:** Direct ptychography upsampling pseudocode

**Inputs:**
- sub-sampled 4D-STEM dataset, $I(\boldsymbol{r}, \boldsymbol{k})$
- (integer) upsampling factor, $f$
- set of aberration coefficients, $C_{n,m}$
- convergence semiangle, $k_0$
- relative rotation angle between scan/detector coordinates, $\theta$

**Output:** ptychographic reconstruction upsampled by a factor $f$

**Initialization:**
- initialize $f$-upsampled reconstruction output, $I'(\boldsymbol{r}') \leftarrow 0$
- define set of BF indices of size $N_{\text{BF}}$, $\boldsymbol{k}_{\text{BF}} = \{\boldsymbol{k} : |\boldsymbol{k}| < k_0\}$
- define set of BF images of size $N_{\text{BF}}$, $I_{\text{BF}} = \{I(\boldsymbol{r}, \boldsymbol{k}) : \boldsymbol{k} \in \boldsymbol{k}_{\text{BF}}\}$
- compute $f$-upsampled and $\theta$-rotated spatial frequency grid, $\boldsymbol{q}'$

**Reconstruction:**

1 **for each** BF index $i = 1$ to $N_{\text{BF}}$:
2  $G \leftarrow \mathcal{F}_{\boldsymbol{r}\rightarrow\boldsymbol{q}}[I_{\text{BF}}[i]]$  (Fourier transformed vBF)
3  $G' \leftarrow \text{tile}_f[G]$  (tile Line 2 to upsampled q-grid)
4  $\boldsymbol{q}'_{\pm} \leftarrow \boldsymbol{q}' \pm \boldsymbol{k}_{\text{BF}}[i]$  (shift q-grid by detector frequency)
5  $\Gamma \leftarrow$ Equation 4  (calculate $\Gamma$ on q-grid using $C_{n,m}$ & $k_0$)
6  $I' += \Im\left[\mathcal{F}^{-1}_{\boldsymbol{q}\rightarrow\boldsymbol{r}'}\left[\frac{G'\Gamma^*}{|\Gamma|}\right]\right]/N_{\text{BF}}$  (filter Line 3 and update output)
7 **end**

| Defocus $\Delta f$ [Å] | (direct) vBF diversity ratio $(\lambda \, |\overline{k_{\mathrm{BF}}}| \, |\Delta f|)/s$ | (iterative) overlap ratio $1 - s/(2 \, |\Delta f| \, \alpha)$ |
|---|---|---|
| 200 | 2.637 | 0.873 |
| 100 | 1.319 | 0.746 |
| 50 | 0.659 | 0.492 |

Table 1: Direct ptychography vBF diversity ratio and iterative ptychography overlap ratio comparison for the defocus values shown in Figure 3.

row), the algorithm successfully recovers the analytical CTF (right column), from 4x sub-sampled acquisitions. As the magnitude of aberrations is reduced (going down), the upsampled reconstructions become noisier. This suggests that the apparent in-plane shifts are smaller than the scan step-size and thus insufficient to recover the information. For defocus-dominated acquisitions, this can be quantified using a vBF diversity rule of thumb:

$$\mathcal{D}_{\mathrm{vBF}} = \frac{\lambda \, |\overline{\boldsymbol{k}_{\mathrm{BF}}}| \, |\Delta f|}{s}, \tag{10}$$

where $s$ is the scan step-size, $\Delta f$ is the defocus, and $|\overline{\boldsymbol{k}_{\mathrm{BF}}}|$ is the mean transverse BF momentum. Successful direct ptychography and parallax imaging (see Supplementary Figure 1) reconstructions require $\mathcal{D}_{\mathrm{vBF}} \gtrsim 1$. Table 1 compares the vBF diversity ratio with the well-known overlap ratio used in iterative ptychography (Gilgenbach et al., 2024), for the defocus values shown in Figure 3 and Supplementary Figure 1. The metrics are in broad agreement, with $\mathcal{D}_{\mathrm{vBF}} \gtrsim 1$ and $\gtrsim 70\%$ overlap for $\Delta f = 100$ Å.

## Low Fluence Performance

The analytical and numerical CTF results presented so far, which suggest that direct ptychography successfully recovers the zero-crossings in the parallax imaging CTF, represent the maximum usable signal. Here, we investigate the performance of the algorithms for Poisson-limited detectors at finite electron fluence.

Figure 4 shows scan-sampled (left) and upsampled (right), parallax imaging (top) and direct imaging (bottom) reconstructions for a simulated array of apoferritin proteins at different electron fluences. At infinite fluence (left column), direct ptychography recovers some of the zero-crossings in the reconstruction FFT. However, as the fluence is decreased (left to right), to realistic values used in imaging biological samples, the two algorithms converge with little-to-no benefit from direct ptychography.

This provides numerical support to the empirical observation that parallax imaging works well at low fluences and is well-suited for imaging beam-sensitive samples (Küçükoğlu et al., 2024, Yu et al., 2024). We also note that imaging of biological samples is usually performed at small convergence angles and strongly defocused; as such, we don't expect higher-order aberrations to introduce significant errors. For physical sciences samples imaged at much higher fluences, we expect both effects (zero crossing recovery and higher-order aberrations) to result in giving direct ptychography an edge over parallax.

## Experimental Reconstructions

Figure 5 demonstrates the utility and limits of direct ptychography upsampling on experimental data of the dose-sensitive metal organic framework (MOF), MOSS-6 (Li et al., 2025). The acquisition uses a very fine step-size of 1.05 Å, and the scan-sampled direct ptychography reconstruction exhibits resolution below 2 nm (top-right panel).

To investigate the upsampling limits of our implementation experimentally, we reconstruct two thinned datasets by taking every second (middle row) and fourth (bottom row) probes. These are shown along the diagonal and, as expected, are band-limited versions of the scan-sampled reconstruction. Impressively, as the thin datasets are computationally upsampled using direct ptychography (left to right), many of the salient features of the scan-sampled reconstruction persist. Note that the bottom right reconstruction uses $1/16^{\mathrm{th}}$ of the original dose, i.e. just 6 e/Å$^2$.

Figure 6 applies the algorithms to publicly-accessible experimental datasets. The twisted $MoS_2$ reconstructions (Jiang et al., 2018), acquired mostly in-focus, demonstrate how i) the parallax imaging CTF suffers in the absence of aberrations (Figure 2, bottom row), and ii) the direct ptychography kernel performs effective denoising by excluding contributions at scattering angles outside the tilt-dependent shifted apertures. The gold nanoparticle reconstructions (Varnavides et al., 2023), demonstrate the power of upsampling sub-sampled acquisitions by revealing lattice information. The Zr-BTB MOF (Li et al., 2025), and virus-like particles datasets (Varnavides et al., 2023), are consistent with the observations in the previous section that, for greatly-defocused acquisitions at low electron fluences, the differences between parallax imaging and direct ptychography are negligible. Note that the $MoS_2$ and MOF datasets were binned in reciprocal-space and thinned in real-space respectively to demonstrate the algorithm robustness and upsampling.

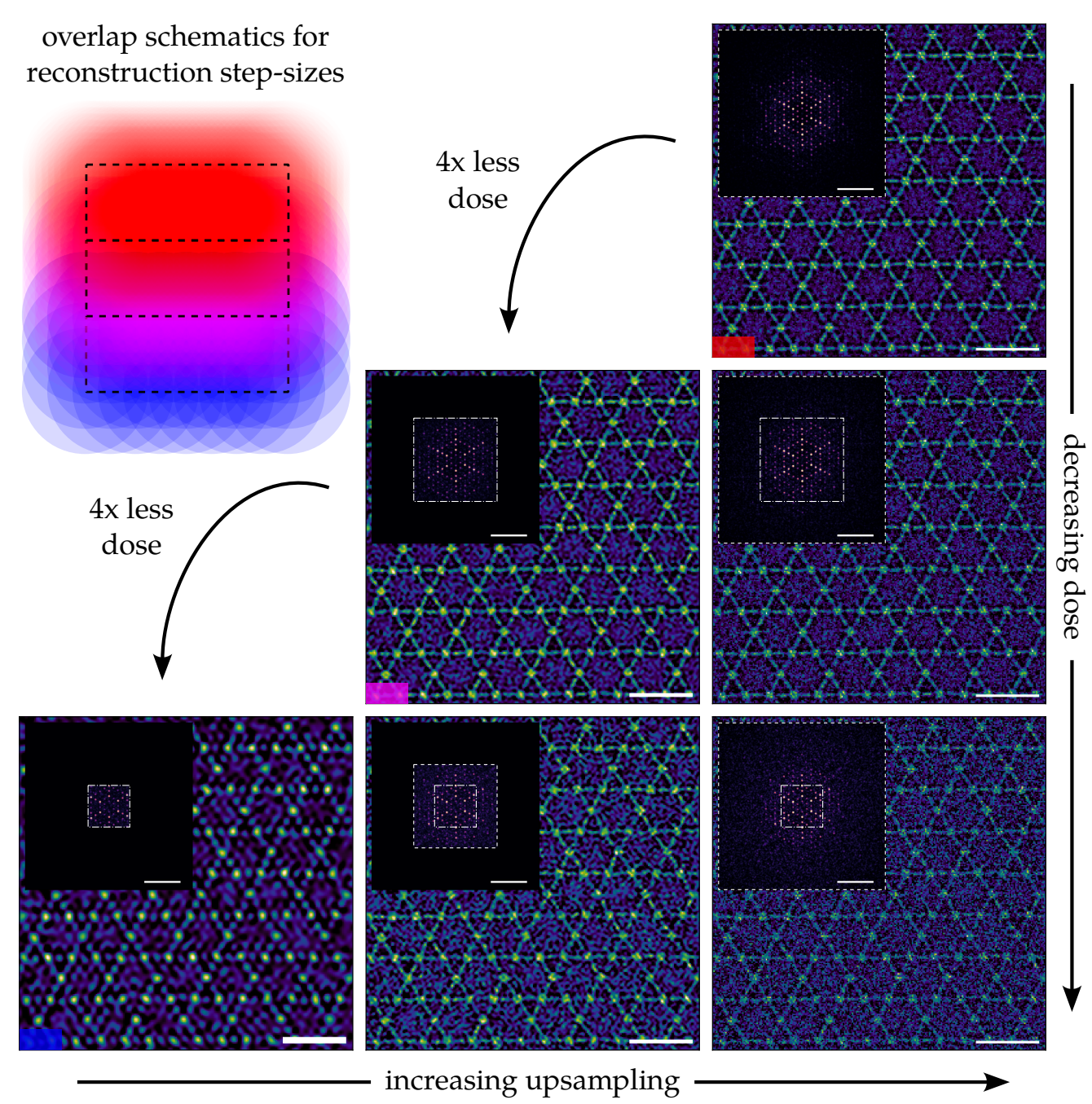

Figure 5: **Experimental upsampling limits.** Direct ptychography reconstruction of a dose-sensitive MOF (top-right). Middle and bottom rows show sub-sampled and upsampled reconstructions by taking every second and fourth probes respectively. Fourier transform insets highlight the scan-sampling (dotted) and upsampling (dashed) reconstruction limits. Real-space and reciprocal-space scalebars are 5 nm and 2 nm$^{-1}$.

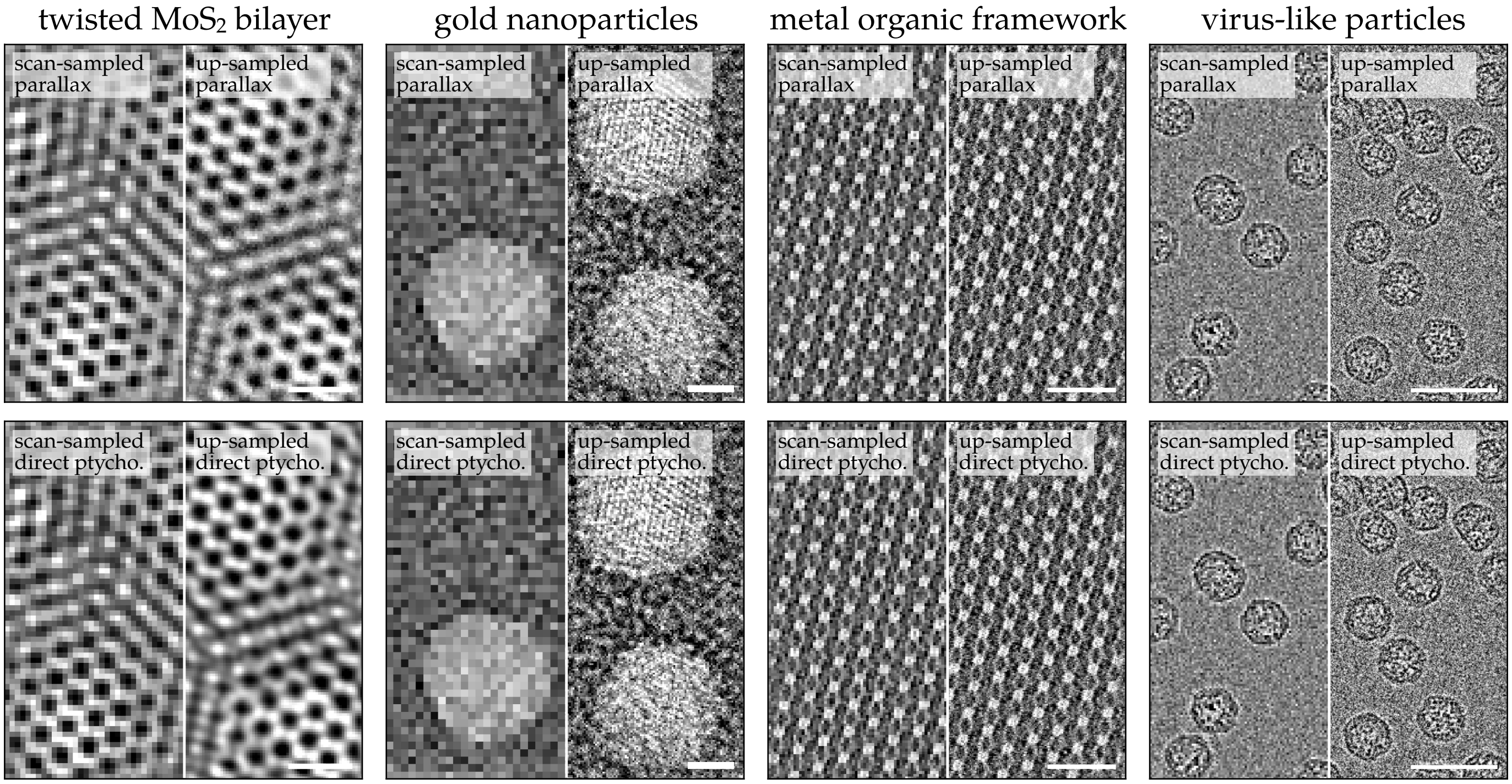

Figure 6: **Upsampled experimental reconstructions.** Scan-sampled (left half) and upsampled (right half), parallax imaging (top row) and direct ptychography (bottom row) reconstructions for various materials classes. Scalebars from left to right are 5 Å, 20 Å, 5 nm, and 50 nm respectively.

As alluded to in Algorithm 1, the experimental reconstructions presented here require accurate knowledge of the low-order aberration coefficients and the relative rotation between the scan and detector geometries. Both direct ptychography (Ning et al., 2024, Yang et al., 2016b), and parallax imaging (Lupini et al., 2016; Varnavides et al., 2024, 2023; Yu et al., 2024), provide self-consistent methods to estimate these from the datasets directly. In our implementation we take a more data-driven approach which works remarkably well for all the datasets we have tried so far. In particular, we optimize the necessary hyper-parameters ($\theta$, $C_{1,0}, C_{1,2}, \varphi_{1,2}$, etc.) against a self-consistent loss function given by the mean variance across the aligned BF images stack (see Supplementary Table 1).

## Conclusions

For weak-phase objects, direct ptychography provides a robust way to accurately recover the sample information. However, most implementations are limited by the scan step size, placing stringent sampling constraints and precluding the use of defocused, large step size acquisitions commonly used for beam sensitive samples. We present an efficient algorithm to overcome this limitation by leveraging the close connection direct ptychography has with parallax imaging – a recently developed phase retrieval technique with upsampling capabilities. The upsampling scheme can further be readily applied on the related techniques optimum bright-field STEM (OBF) (Ooe et al., 2024), and single-sideband integrated center-of-mass imaging (SSB-ICOM) (Ding et al., 2025).

We investigate the transfer of information of these two techniques, the limits of the upsampling scheme, and the dose efficiency of the algorithms in simulated and experimental datasets. We find that, for samples imaged at high electron fluence or for acquisitions dominated by higher-order aberrations, direct ptychography is expected to outperform parallax imaging – recovering the latter's CTF zero crossings. We expect these conditions to be met for physical sciences samples imaged at high resolution, and expect its low computational cost and relaxed sampling requirements to make it attractive to STEM practitioners in the future. For dose-sensitive samples, especially organic and biological samples usually imaged with smaller convergence angles and dominated by defocus, we find that parallax imaging provides a robust reconstruction. We hope this exposition of parallax imaging – as a quadratic approximation to direct ptychography – will place it on stronger theoretical footing and encourage its wider spread adoption.

## Availability of Data and Materials

The datasets/notebooks to reproduce the CTF results are available at https://doi.org/10.69761/ehch7395. The raw experimental datasets, are obtained from the respective references in Varnavides et al. (2023), Jiang et al. (2018), and Li et al. (2025). All other processed datasets/notebooks are available at [ToDo: Add Zenodo link]. The parallax imaging and direct ptychography open-source implementations are available in the quantem package.

## Author Contributions Statement

G.V., J.M.B.: conceptualization, data curation, formal analysis, investigation, methodology, software, validation, visualization, writing – original draft, writing – review & editing; S.M.R.: validation, writing - review & editing; M.C.S, L.J., C.O.: funding acquisition, project administration, supervision, writing - review & editing; All authors read and approved the final version of the manuscript.

## Acknowledgements

The authors gratefully acknowledge Dr. Colum O'Leary, Dr. Steven E Zeltmann, Dr. Andrew Lupini, Prof. David A Muller, and Prof. Peter Nellist for helpful discussions.

## Financial Support

Work at the Molecular Foundry was supported by the Office of Science, Office of Basic Energy Sciences, of the U.S. Department of Energy under Contract No. DE-AC02-05CH11231. J.M.B. and L.J. acknowledge support from Research Ireland grant number 19/FFP/6813. L.J. acknowledges support from Royal Society and Research Ireland grant numbers URF/RI/191637 and 12/RC/2278_P2. S.M.R. and C.O. acknowledge support from the DOE Early Career Research Program.

## Conflict of Interest

The authors declare there are no conflicts of interest.

| **Experimental dataset** | $\theta$ [°] | $C_{1,0}$ [nm] | $C_{1,2}$ [nm] | $\varphi_{1,2}$ [°] |
|---|---|---|---|---|
| MOSS-6 (Figure 5) | 94.6 | −108.6 | 11.4 | −33.6 |
| Zr-BTB (Figure 6) | 91.2 | −51.0 | 40.3 | 6.0 |
| gold nanoparticles | 168.8 | −40.8 | 0.5 | −25.4 |
| twisted $MoS_2$ | 29.9 | −6.3 | 1.1 | 73.8 |
| virus-like particles | 5.2 | 1670 | 13.7 | 23.1 |

Supplementary Table 1: Optimized hyper-parameters for experimental reconstructions shown in Figure 5 and Figure 6.

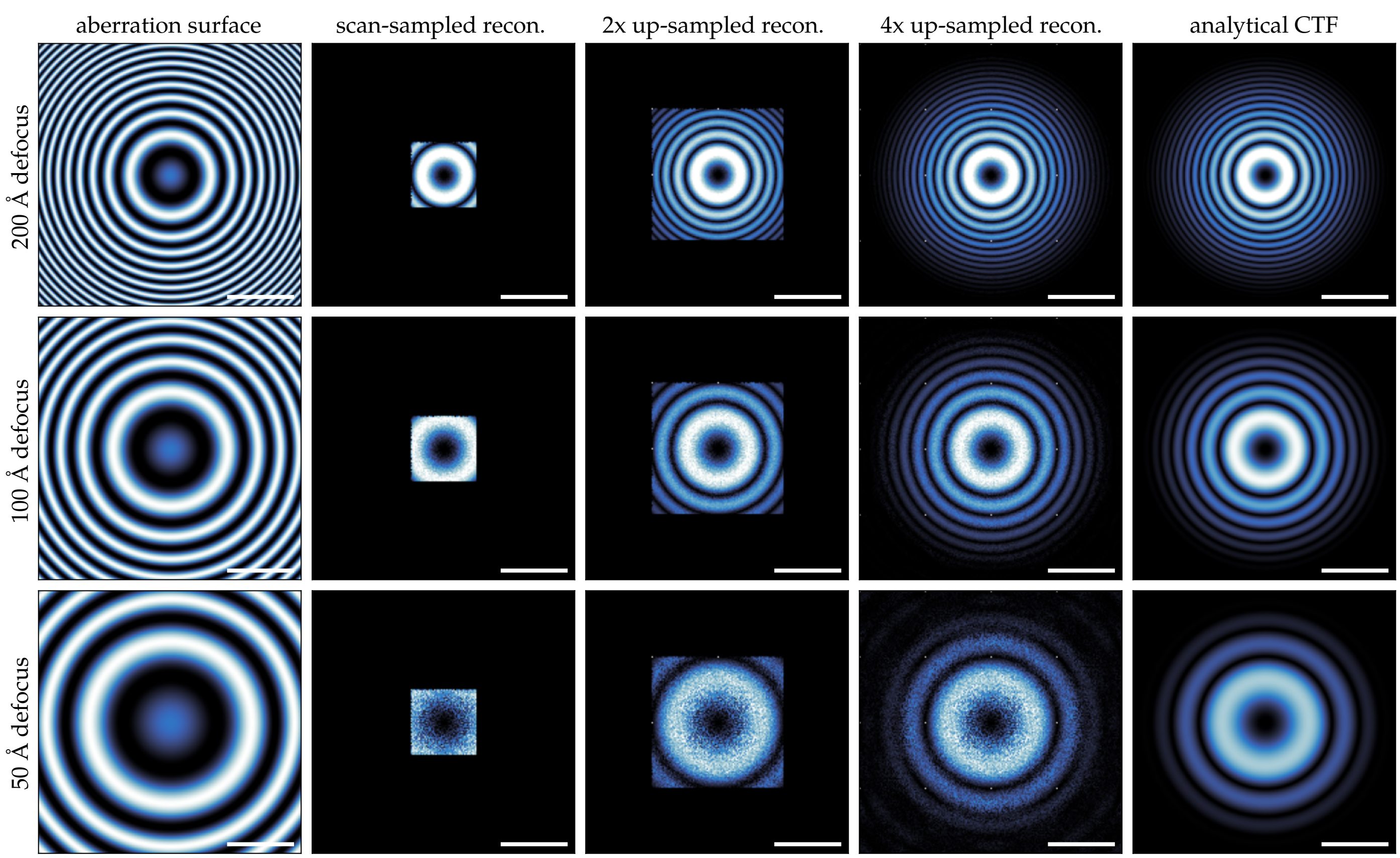

Supplementary Figure 1: **Upsampling limits in parallax imaging.** Subsampled parallax acquisitions (second column), upsampled to 2x (third column) and 4x (fourth column) scan resolution. The last column plots the analytical CTF the upsampled reconstructions are attempting to recover. Each row represents an increasing amount of defocus, highlighting upsampling factor limits are proportional to the aberrations magnitude. Scalebars are $k_0$.